\documentclass[aps,preprint,floats,nofootinbib]{revtex4}
\usepackage{amsmath}
\usepackage{amssymb}
\usepackage{latexsym}
\usepackage{epsf}
\usepackage{graphicx}

\newlength{\defbaselineskip}
\newcommand{\setlinespacing}[1]%
           {\setlength{\baselineskip}{#1 \defbaselineskip}}

\def\lsim{\mathrel{\raise.3ex\hbox{$<$\kern-.75em\lower1ex\hbox{$\sim$}}}} 
\def\gsim{\mathrel{\raise.3ex\hbox{$>$\kern-.75em\lower1ex\hbox{$\sim$}}}} 
\begin{document}

\preprint{
\hfill
\begin{minipage}[t]{3in}
\begin{flushright}
\vspace{0.0in}
FERMILAB--PUB--08--296--A 
\end{flushright}
\end{minipage}
}

\hfill$\vcenter{\hbox{}}$

\vskip 0.5cm

\title{The New DAMA Dark-Matter Window and Energetic-Neutrino Searches}
\author{Dan Hooper$^1$, Frank Petriello$^2$, Kathryn M.
Zurek$^2$, and Marc Kamionkowski$^3$}
\address{$^1$Theoretical Astrophysics, Fermi National
Accelerator Laboratory, Batavia, IL  60510; 
$^2$Physics Department, University of Wisconsin, Madison, WI
53706;
$^3$California
Institute of Technology, Mail Code 130-33, Pasadena, CA 91125
}

\date{\today}

\bigskip

\begin{abstract}

Recently, the DAMA/LIBRA collaboration has repeated and
reinforced their claim to have detected an annual modulation in
their signal rate, and have interpreted this observation as
evidence for dark-matter particles at the 8.2$\sigma$ confidence
level. Furthermore, it has also been noted that the effects of
channeling may enable a WIMP that scatters elastically via
spin-independent interactions from nuclei to produce
the signal observed by DAMA/LIBRA without exceeding the limits placed by CDMS, XENON, CRESST, CoGeNT and other direct-detection experiments. 
To accommodate this signal, however, the mass of the responsible
dark-matter particle must be relatively light,  $m_{\rm DM} \lsim 10$ GeV. 
%Over the range of masses and elastic scattering cross sections
%consistent with DAMA,
Such dark-matter particles will become captured by and
annihilate in the Sun at very high rates, leading to a
potentially large flux of GeV-scale neutrinos. We calculate the
neutrino spectrum resulting from WIMP annihilations in the Sun
and show that existing limits from Super-Kamiokande can be used
to close a significant portion of the DAMA region, especially if the
dark-matter particles produce tau leptons or neutrinos in a
sizable fraction of their annihilations. We also determine the
spin-dependent WIMP-nuclei elastic-scattering parameter space consistent
with DAMA.  The constraints from Super-Kamiokande on the
spin-dependent scenario are even more severe---they exclude any
self-annihilating WIMP in the DAMA region that annihilates $1\%$
of the time or more to any combination of neutrinos, tau
leptons, or charm or bottom quarks.

\end{abstract}

%\pacs{PAC numbers: 95.35.+d; 95.85.Ry; 11.30.Pb; 04.50.+h}
\maketitle

\section{Introduction}

Recently, the DAMA collaboration has provided further evidence for an annual modulation in the rate of nuclear-recoil events observed in their experiment~\cite{Bernabei:2008yi}.  Such a signal arises naturally from postulating weakly interacting massive particles (WIMPs) in the Galactic halo that scatter from target nuclei in detectors.  The annual modulation of the interaction rate results from the variation in the relative velocity of the Earth with respect to the Galactic dark-matter halo as the Earth orbits the Sun.  This changes the flux of dark-matter particles and their velocity distribution, with expected extrema occurring at June 2 and December 2.  The DAMA experiment 
observes a maximum rate at low nuclear-recoil energies on May 24, plus or minus 8 days, and have accumulated enough data to put the significance of the observed modulation at approximately $8\sigma$.   Both the phase and amplitude of the signal are highly suggestive of WIMP interactions.  The collaboration has not been able to 
identify any other systematic effects capable of producing this signal, and have claimed that the annual modulation is a discovery of dark matter.  This claim has been 
controversial, partly because a number of other experiments appear to be in direct contradiction.  

Several studies have attempted to reconcile the DAMA modulation signal with the null results of other direct-detection experiments~\cite{Bottino1,Foot:2008nw,Feng:2008dz,
Petriello:2008jj,Bottino:2008mf,Chang:2008gd}, assuming a
spin-independent WIMP-nucleus interaction.  A feature common to the analyses of Refs.~\cite{Bottino1,Feng:2008dz,Petriello:2008jj,Bottino:2008mf,Chang:2008xa,Fairbairn:2008gz} is that an elastically scattering WIMP with a mass in the several GeV range can satisfy the results of DAMA.   The consistency of this region with the null results of CDMS~\cite{cdms}, CRESST~\cite{cresst}, CoGeNT~\cite{cogent}, and XENON~\cite{xenon} depends on the details of how the DAMA recoil energy spectrum is fit.  If the data is divided into the two bins $2-6\;{\rm keVee}$ and 
$6-14\;{\rm keVee}$, where keVee is the electron-equivalent recoil energy, then DAMA and the null experiments can be simultaneously accommodated by $3-8\;{\rm GeV}$ WIMPs.  If the DAMA modulation data is binned more 
finely, then the modulation in the $2-2.5\;{\rm keVee}$ bin
right at the DAMA threshold is difficult to reconcile with other
direct-detection constraints.  We refer the reader to the studies in 
Refs.~\cite{Petriello:2008jj,Chang:2008xa,Fairbairn:2008gz} for further details.  The allowed parameter region depends crucially on the occurrence of channeling in the NaI crystals of the DAMA apparatus, an effect noted in Ref.~\cite{Drobyshevski:2007zj} and studied by the DAMA collaboration~\cite{Bernabei:2007hw}. 

In this paper, we consider the constraints to the new DAMA
dark-matter parameter space that come from null searches for
energetic neutrinos from the Sun~\cite{neutrino}.  If WIMPs scatter from
nuclei in DAMA, then they will also scatter from nuclei in the
Sun, be captured and thus annihilate therein, and thus produce
energetic neutrinos that can be sought in experiments 
such as IceCube~\cite{icecube}, AMANDA~\cite{amanda},
Baksan~\cite{baksan}, MACRO~\cite{macro},
ANTARES~\cite{antares}, and Super-Kamiokande~\cite{superk}.  The
only caveat is that the energetic-neutrino spectra will depend
on the WIMP-annihiation products.  Still, the range of neutrino
spectra is bracketed,
and so model-independent bounds can be sometimes
obtained~\cite{Kamionkowski:1994dp}.  In particular, the Sun is
composed primarily of protons (nuclei with spin), and so null
neutrino searches should be especially constraining for an
explanation of DAMA in terms of a WIMP with spin-{\it de}pendent
interactions, as shown in Ref.~\cite{Ullio:2000bv} for the old
spin-dependent DAMA parameter space.

Here, we determine the regions of the mass--cross-section parameter
space implied by the new DAMA results for a spin-dependent WIMP,
and we determine
parameter space (for both spin-dependent and spin-independent
WIMPs) eliminated by null neutrino searches.  In order to extend
the analysis of Ref.~\cite{Ullio:2000bv} to the lower WIMP
masses implied by channeling, we calculate the full neutrino
energy spectra produced by decays of tau leptons and charm and
bottom quarks, and we consider prompt annihilation to neutrinos.
We also include the effects of neutrino mixing and the effects
of WIMP evaporation from the Sun.
The muon energy thresholds of ANTARES and IceCube are 10 GeV or
higher and are therefore not able to observe the neutrinos
produced by a several-GeV WIMP.  We therefore consider bounds
arising from the Super-Kamiokande experiment, which can identify
muons with energies as low as 1.6 GeV.  The limits imposed by
Baksan and MACRO are similar to those coming form
Super-Kamiokande.  
We study the DAMA allowed region arising from both the
two-bin study of Ref.~\cite{Petriello:2008jj} and the parameter  
space arising from the full spectral analysis both with and
without the $2-2.5\;{\rm keVee}$ bin~\cite{Chang:2008xa,Fairbairn:2008gz}.

Our conclusions are that that upper limits to the flux of energetic
neutrinos from the Sun severely constrain the DAMA spin-independent
parameter space if WIMPs annihilate directly to neutrinos or to
tau leptons.  The constraints from
Super-Kamiokande on the spin-dependent scenario are even more
severe; they exclude any spin-dependent WIMP in the DAMA region that
annihilates $1\%$ of the time or more to any combination of
neutrinos, tau leptons, or charm or bottom quarks.

Our paper is organized as follows.  We review in
Section~\ref{sec:formalism} the formalism of WIMP capture and
annihilation in the Sun, including the effect of evaporation,
which is important for dark matter near the 3-GeV lower edge of
the DAMA allowed region.  In Section~\ref{sec:limits} we discuss
the detection of the neutrinos from WIMP annihilation using
upward-going muons in the
Super-Kamiokande detector and study the constraints imposed by
this process on the DAMA allowed region.  We also present in
Sec.~\ref{sec:limits} the spin-dependent parameter space
(including channeling) implied by DAMA.  In Sec.~\ref{susy}, we
comment on the energetic-neutrino constraint as applied to the case of light neutralino dark matter. Finally, we summarize our conclusions in Section~\ref{sec:conc}.

\section{WIMP Capture and Annihilation in the Sun \label{sec:formalism}}

We briefly review here the basic formulae describing WIMP capture and annihilation in the Sun.  A generic species of dark-matter particle present in the solar system will scatter elastically with and become captured in the Sun at a rate given by~\cite{capture}
\begin{eqnarray}
C^{\odot} \simeq 1.3 \times 10^{25} \, \mathrm{s}^{-1} 
\left( \frac{\rho_{\rm DM}}{0.3\, \mathrm{GeV}/\mathrm{cm}^3} \right) 
\left( \frac{270\, \mathrm{km/s}}{\bar{v}} \right)  
\left( \frac{1 \, \mathrm{GeV}}{m_{\rm DM}} \right)  \nonumber \\
\times \bigg[ \bigg(\frac{\sigma_{\mathrm{H}}}{10^{-40}\, {\rm cm}^2}\bigg)  S(m_{\rm DM}/m_{{\rm H}}) +  1.1 \bigg(\frac{\sigma_{\mathrm{He}}}{16 \times 10^{-40}\, {\rm cm}^2}\bigg)  S(m_{\rm DM}/m_{{\rm He}}) \bigg],
\label{capture}
\end{eqnarray}
where $\rho_{\rm DM}$ is the local dark-matter density, $\bar{v}$ is the local root-mean-square velocity of halo dark-matter particles, and $m_{\rm DM}$ is the dark-matter mass. $\sigma_{\mathrm{H}}$ and $\sigma_{\mathrm{He}}$ are the elastic scattering cross sections of the WIMP with hydrogen and helium nuclei, respectively. The factor of $1.1$ reflects the solar abundance of helium relative to hydrogen and well as dynamical factors and form factor suppression. The quantity $S$ is a kinetic suppression factor given by
\begin{equation}
S(x)=\bigg[\frac{A(x)^{3/2}}{(1+A(x)^{3/2})}\bigg]^{2/3},
\end{equation}
where
\begin{equation}
A(x)=\frac{3x}{2(x-1)^2}\bigg(\frac{\langle v_{\rm esc}
\rangle}{\bar{v}}\bigg)^2.
\end{equation}
For WIMPs much heavier than hydrogen and helium nuclei, this suppression can be considerable. For WIMPs in the 1--10 GeV range, however, $S \approx 0.9-1.0$.

If the capture rate and annihilation cross sections are sufficiently large, equilibrium will be reached between these processes.   The differential equation 
governing the number of WIMPs in the Sun, denoted here as $N$, is
\begin{equation}
\dot{N} = C^{\odot} - A^{\odot} N^2 - E^{\odot} N,
\end{equation}
where $C^{\odot}$ is the capture rate, $A^{\odot}$ is the 
annihilation cross section times the relative WIMP velocity per unit volume, and $E^{\odot}$ is the inverse time for a WIMP to exit the Sun via evaporation. $A^{\odot}$ can be approximated by
\begin{equation}
A^{\odot} = \frac{\langle \sigma v \rangle}{V_{\mathrm{eff}}}, 
\end{equation}
where $V_{\mathrm{eff}}$ is the effective volume of the core
of the Sun determined roughly by matching the core temperature with 
the gravitational potential energy of a single WIMP at the core
radius.  This was found in Refs.~\cite{equ1,equ2} to be
\begin{equation}
V_{\rm eff} = 5.7 \times 10^{30} \, \mathrm{cm}^3 
\left( \frac{1 \, \mathrm{GeV}}{m_{\rm DM}} \right)^{3/2} \;.
\end{equation}
Neglecting evaporation for the moment, the present WIMP annihilation rate is given by
\begin{equation} 
\Gamma = \frac{1}{2} A^{\odot} N^2 = \frac{1}{2} \, C^{\odot} \, 
\tanh^2 \left( t_{\odot}/\tau_E \right) \;, 
\label{rate:noevap}
\end{equation}
where $t_{\odot} \simeq 4.5$ billion years is the age of the solar system, and $\tau_E =\left(C^{\odot} A^{\odot}\right)^{-1/2}$ is the time required to reach equilibrium.
The annihilation rate is maximized when it reaches equilibrium with
the capture rate.  This occurs when 
\begin{equation}
t_{\odot}/\tau_E \gg 1 \; .
\end{equation}
In the case in question, this condition will be met so long as
\begin{equation}
\langle \sigma v \rangle \gsim 3 \times 10^{-30} \, {\rm cm}^3/{\rm s} \, \bigg(\frac{1 \, {\rm GeV}}{m_{\rm DM}}\bigg)^{1/2} \, \bigg(\frac{10^{-40}\, {\rm cm}^2}{\sigma_{\mathrm{H}}}\bigg). 
\end{equation}
Once equilibrium is reached, the final annihilation rate (and corresponding neutrino flux and event rate) has no further dependence on the dark-matter particle's annihilation cross section.

\begin{figure}
\resizebox{11.2cm}{!}{\includegraphics{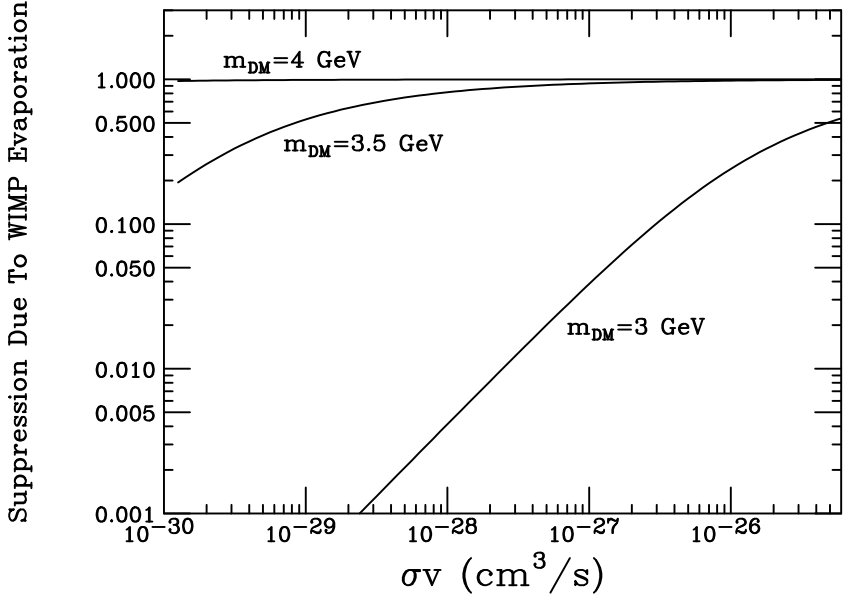}}
\caption{The factor by which the WIMP annihilation rate in the Sun is suppressed as a result of WIMP evaporation. For each WIMP mass, we used a spin-independent elastic scattering cross section near the middle of the DAMA region (see the upper frame of Fig.~\ref{limitSI}). For WIMPs heavier than 4 GeV, the effect of evaporation is negligible.}
\label{evapor}
%\end{center}
\end{figure}

For WIMPs with masses in the range being considered here, the process of WIMP evaporation from the Sun could also be potentially important. Ref.~\cite{evaporation} found an approximate timescale for WIMP evaporation given by (see also Ref.~\cite{Griest:1986yu}):
\begin{equation}
E^{\odot} \approx 10^{-(\frac{7}{2} (m_{\rm DM}/{\rm GeV}) + 4)} \, {\rm s}^{-1} \, \bigg(\frac{\sigma_{\rm H}}{5 \times 10^{-39}\, {\rm cm}^2}\bigg).
\end{equation}
The WIMP annihilation rate in the presence of evaporation is~\cite{Griest:1986yu}
\begin{equation}
\Gamma = \frac{1}{2} \, C^{\odot} \, \left[ \frac{\tanh \left(\alpha  t_{\odot}/\tau_E \right)}{\alpha+\frac{1}{2}E^{\odot}\tau_E \tanh \left(\alpha  t_{\odot}/\tau_E \right)} \right]^2,
\label{rate:evap}
\end{equation}
with $\alpha=\left\{1+(E^{\odot}\tau_E/2)^2\right\}^{1/2}$.  For the lightest range of WIMP masses in the DAMA region (3--4 GeV), the timescale for the process of WIMP evaporation can be comparable to the time required to reach capture-annihilation equilibrium, $E^{\odot} \tau_E \gsim 1$, and thus may potentially reduce the annihilation rate.  In Fig.~\ref{evapor}, we show the suppression of WIMP annihilation resulting from evaporation as a function of the WIMP's annihilation cross section, obtained from the ratio of Eq.~(\ref{rate:evap}) and 
Eq.~(\ref{rate:noevap}).  For WIMPs heavier than $\sim 4$ GeV, evaporation plays a negligible role. For lighter WIMPs, however, the annihilation rate and corresponding neutrino flux may be suppressed, depending on the magnitude of the annihilation cross section. In our calculations, we will use a dark-matter annihilation cross section of $\sigma v = 3 \times 10^{-26}$ cm$^3$/s, which is generally expected for a thermal relic in the absence of resonant annihilation, coannihilations or strong $s$-wave suppression. If the annihilation cross section is smaller than this value, the effects of WIMP evaporation on the annihilation rate and corresponding neutrino flux may be more pronounced, as shown in Fig.~\ref{evapor}.

\section{Limits from muon production in Super-Kamiokande \label{sec:limits}}
 
As the WIMPs annihilate, they can generate neutrinos through a wide range of channels. Annihilations to bottom quarks, charm quarks, and tau leptons each generate neutrinos in their subsequent decays. In some models, WIMPs can also annihilate directly to neutrino pairs. Once produced, neutrinos travel to the Earth where they can be detected.  The spectra of muon neutrinos and antineutrinos at the Earth from WIMP annihilations in the Sun is given by
\begin{eqnarray}
\label{wimpflux1}
\frac{dN_{\nu_{\mu}}}{dE_{\nu_{\mu}}} &\approx& \frac{1}{3}\frac{ C_{\odot} F_{\rm{Eq}}}{4 \pi D_{\rm{ES}}^2}   \bigg(\frac{dN_{\nu_e}}{dE_{\nu_e}}+\frac{dN_{\nu_{\mu}}}{dE_{\nu_{\mu}}}+\frac{dN_{\nu_{\tau}}}{dE_{\nu_{\tau}}}\bigg)^{\rm{Inj}},\\
\frac{dN_{\bar{\nu}_{\mu}}}{dE_{\bar{\nu}_{\mu}}} &\approx& \frac{1}{2}\frac{ C_{\odot} F_{\rm{Eq}}}{4 \pi D_{\rm{ES}}^2}   \bigg(\frac{dN_{\bar{\nu}_{\mu}}}{dE_{\bar{\nu}_{\mu}}}+\frac{dN_{\bar{\nu}_{\tau}}}{dE_{\bar{\nu}_{\tau}}}\bigg)^{\rm{Inj}},
\label{wimpflux2}
\end{eqnarray}
where $C_{\odot}$ is the WIMP capture rate in the Sun, $F_{\rm{Eq}}$ is the non-equilibrium suppression factor ($\approx 1$ for capture-annihilation equilibrium), $D_{\rm{ES}}$ is the Earth-Sun distance and the bracketed quantities are the neutrino and antineutrino spectra from the Sun per annihilating WIMP. Due to $\nu_{\mu}-\nu_{\tau}$ vacuum oscillations and MSW enhanced $\nu_e$ oscillations~\cite{msw} in the Sun, the muon neutrino flux is approximately given by the average of the $\nu_{\mu}$, $\nu_{e}$ and $\nu_{\tau}$ components, leading to the factor of $1/3$ in Eq.~(\ref{wimpflux1}). The oscillations of anti-electron neutrinos are MSW suppressed, however, leading to a flux of anti-muon neutrinos which is the average of the $\bar{\nu}_{\mu}$ and $\bar{\nu}_{\tau}$ components~\cite{Cirelli:2005gh}.

Muon neutrinos produce muons in charged-current interactions with nuclei inside of and around the detector volume of Super-Kamiokande. In the analysis of the Super-Kamiokande collaboration~\cite{superk}, upward-going muon tracks extending 7 meters or more within the inner detector were counted.  Such events are produced through dark-matter annihilations in the Sun at a rate given by
\begin{eqnarray}
R_{\rm{events}} &\approx& \int \int \frac{1}{2}\frac{dN_{\nu_{\mu}}}{dE_{\nu_{\mu}}}\, \frac{d\sigma_{\nu}}{dy}(E_{\nu_{\mu}},y) \, R_{\mu}(E_{\mu})  A_{\rm{Det}} \, N_A \, dE_{\nu_{\mu}} \, dy \nonumber \\
 &+& \int \int \frac{1}{2}\frac{dN_{\bar{\nu}_{\mu}}}{dE_{\bar{\nu}_{\mu}}}\, \frac{d\sigma_{\bar{\nu}}}{dy}(E_{\bar{\nu}_{\mu}},y) \, R_{\mu}(E_{\mu})  A_{\rm{Det}} \, N_A \, dE_{\bar{\nu}_{\mu}} \, dy.
\end{eqnarray}
Here, $\sigma_{\nu}(E_{\nu_{\mu}}) \approx 8 \times 10^{-39} \, {\rm cm}^2 \times [E_{\nu}/({\rm GeV})]$ and $\sigma_{\bar{\nu}}(E_{\bar{\nu}_{\mu}}) \approx 3 \times 10^{-39} \, {\rm cm}^2 \times [E_{\bar{\nu}}/({\rm GeV})]$ are the neutrino-nucleon and antineutrino-nucleon charged-current interaction cross sections, $(1-y)$ is the fraction of the neutrino's (or antineutrino's) energy which goes into the muon, $A_{\rm{Det}}$ is the effective area of the detector, $N_A$ is Avogadro's number, which for water is the number density of protons, and $R_{\mu}(E_{\mu})$ is the range a muon travels before losing its energy. For the dimensions of the inner detector, we take $A_{\rm{Det}}=900$ m$^2$ and a height of 36.2 meters, for a total target volume of 32,500 m$^3$, or 32.5 metric tons. To account for the 7-meter cut applied in the Super-Kamiokande analysis, we substitute the physical muon range ($R_{\mu} \approx 5$ meters $\times E_{\mu}/{\rm GeV}$), with zero if $R_{\mu} < 7$ meters or otherwise with $R_{\mu}+(36.2-7)$ meters. The factor of 1/2 accounts for the Sun being below the Super-Kamiokande detector approximately 50\% of the time.

For the injection spectrum of neutrinos produced in WIMP annihilations, we consider the following annihilation channels: $\nu \bar{\nu}$, $\tau^+ \tau^-$, $c\bar{c}$, and $b \bar{b}$. We neglect annihilation products such as muon pairs, light mesons, etc. as they are expected to lose the vast majority of their energy before decaying and thus do not produce energetic neutrinos. For the case of annihilations to neutrino-antineutrino pairs, we have assumed that equal quantities of each flavor are produced. If instead all of the neutrinos and antineutrinos produced were of electron (muon or tau) flavor, the rate would be 27\% smaller (14\% larger).

For annihilations to tau pairs, we include the semi-leptonic decays $\tau \rightarrow \mu \nu \nu$, $e \nu \nu$, as well as from the hadronic decays $\tau \rightarrow \pi \nu$, $K \nu$, $\pi \pi \nu$, and $\pi \pi \pi \nu$. For charm and bottom quarks, only semileptonic decays contribute.  We model the neutrino energy spectra coming from these decays using updated versions of the 
formulae in Ref.~\cite{Jungman:1994jr} including tau and electron neutrinos.  For the hadronic $\tau$ decays we use approximate formulae that reproduce reasonably well the results 
obtaining using HERWIG~\cite{Grellscheid:2007tt}.  We neglect the possible production of quarkonia near the $b\bar{b}$ threshold and instead use the results for open-flavor semi-leptonic 
decay; as the neutrinos from $b$-decay do not impose strong constraints we believe this approximation is sufficient.

\begin{figure}[htbp]
%\begin{center}

 \includegraphics[width=0.65\textwidth,angle=0]{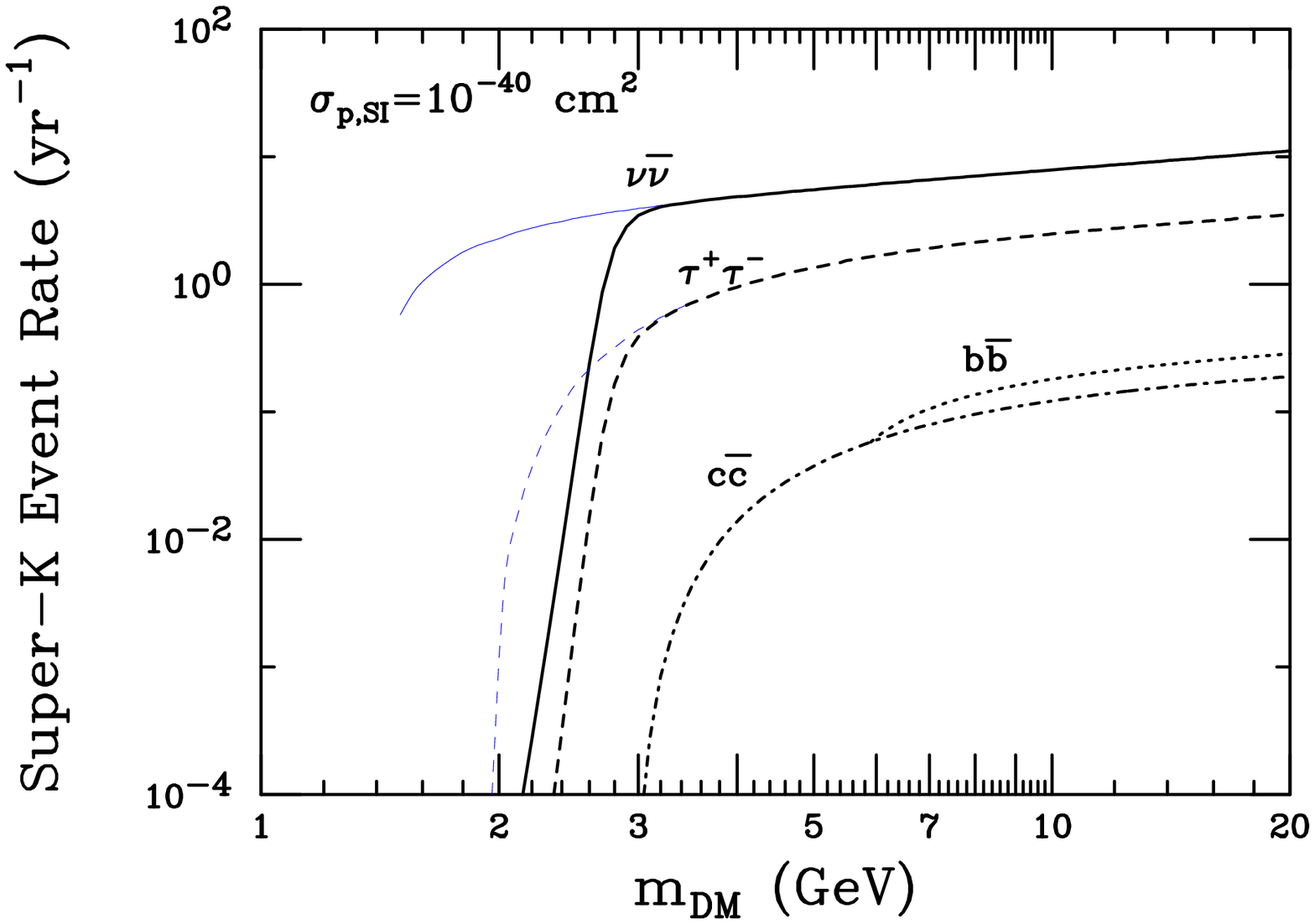}
 \includegraphics[width=0.65\textwidth,angle=0]{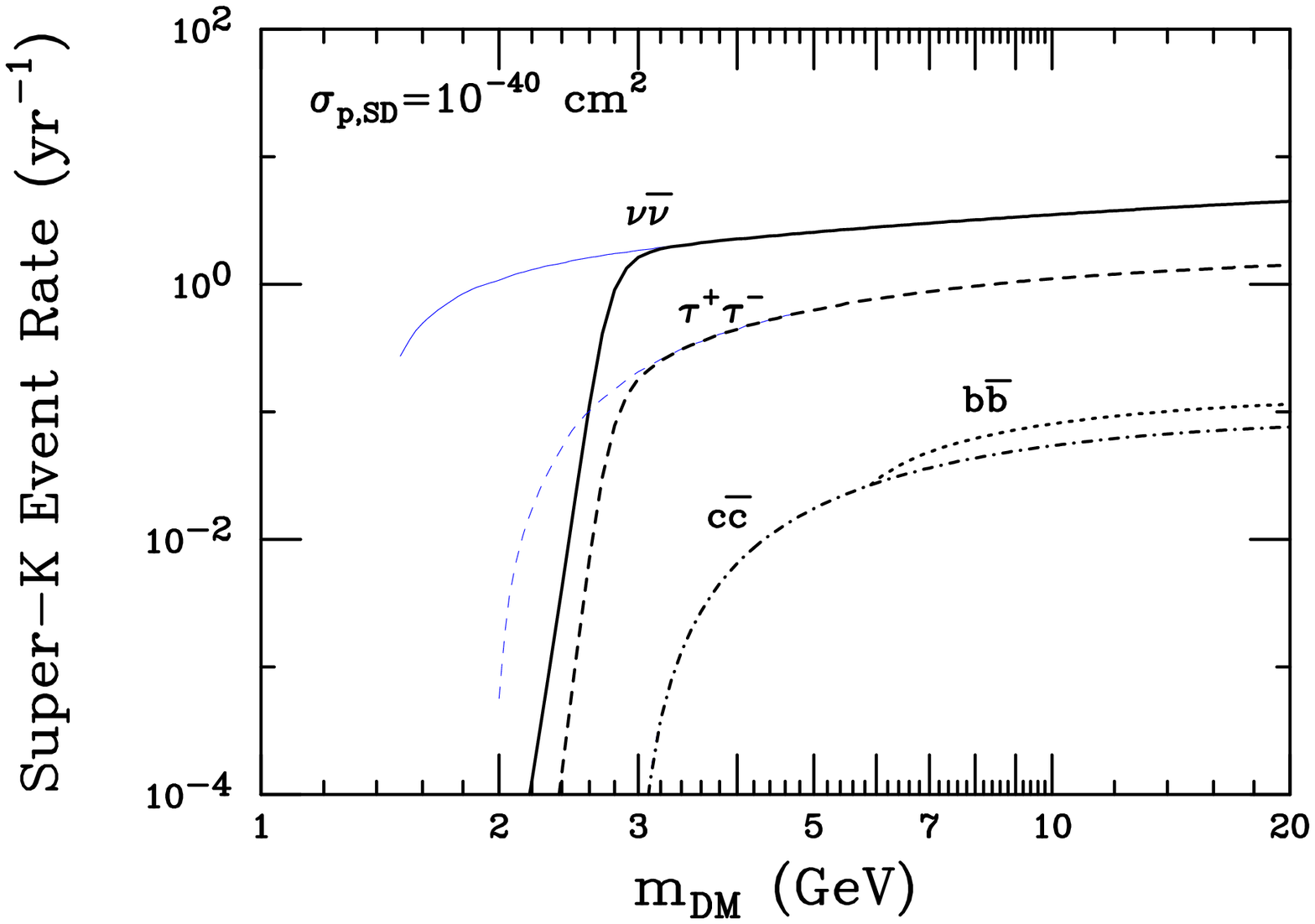}

%\resizebox{11.2cm}{!}{\includegraphics{sunrateSI.ps}}
%\resizebox{11.2cm}{!}{\includegraphics{sunrateSD.ps}}
\caption{The number of upward-going muon events per year in Super-Kamiokande from WIMPs annihilating in the Sun as a function of mass for an elastic scattering cross section with protons of $10^{-40}$ cm$^2$, assuming 100\% annihilation to the indicated channel. The upper and lower frames correspond to spin-independent and spin-dependent couplings, respectively. In each frame, the thin (blue) lines extending to the left denote the results neglecting the effects of WIMP evaporation. See text for more details.}
\label{rate}
%\end{center}
\end{figure}

In Fig.~\ref{rate}, we plot the rate of events at
Super-Kamiokande predicted from annihilating WIMPs. In the upper
frame, we consider spin-independent elastic scattering with
nuclei, which in Eq.~(\ref{capture}) corresponds to $\sigma_{\rm
H}=\sigma_{p,SI}$ and $\sigma_{\rm He}=16 \, \sigma_{p,SI}$. In
the lower frame, we consider spin-dependent scattering with
$\sigma_{\rm H}=\sigma_{p,SD}$ and $\sigma_{\rm He}=0$ (we
neglect the tiny contribution from spin-dependent scattering
from $^3$He). For comparison, we also show as thin (blue) lines
in these frames the rate neglecting the effects of WIMP
evaporation. 

% Before continuing, we note one caveat regarding the spin-dependent scattering cross section with helium.  This is not strictly zero since the Sun contains a tiny component of $^{3}{\rm He}$, measured to have an isotopic abundance of roughly $2 \times 10^{-4}$ in the local interstellar cloud~\cite{hesun}.  Although the amount of $^3$He in the core of the Sun could be greater, it is unlikely 
% to be smaller by a very significant amount.  Including all factors, we estimate the spin-dependent cross section with helium to be $10^{-5}$ times the cross section with hydrogen.  Since it 
% is difficult to imagine a particle physics model that could suppress the scattering from helium relative to hydrogen by this amount, we neglect this correction in our analysis.

\begin{figure}[htbp]
%\begin{center}

 \includegraphics[width=0.63\textwidth,angle=0]{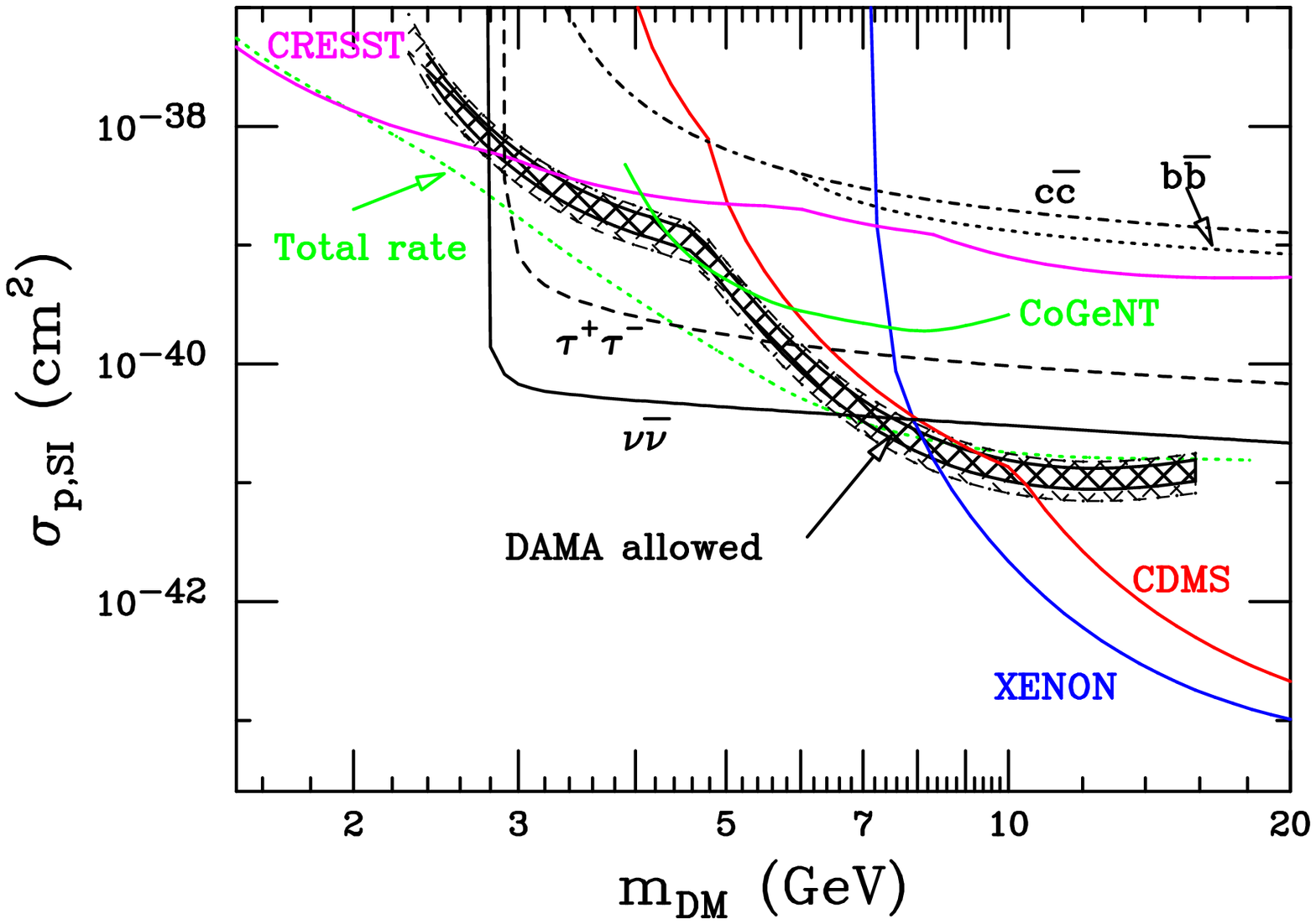}
 \includegraphics[width=0.63\textwidth,angle=0]{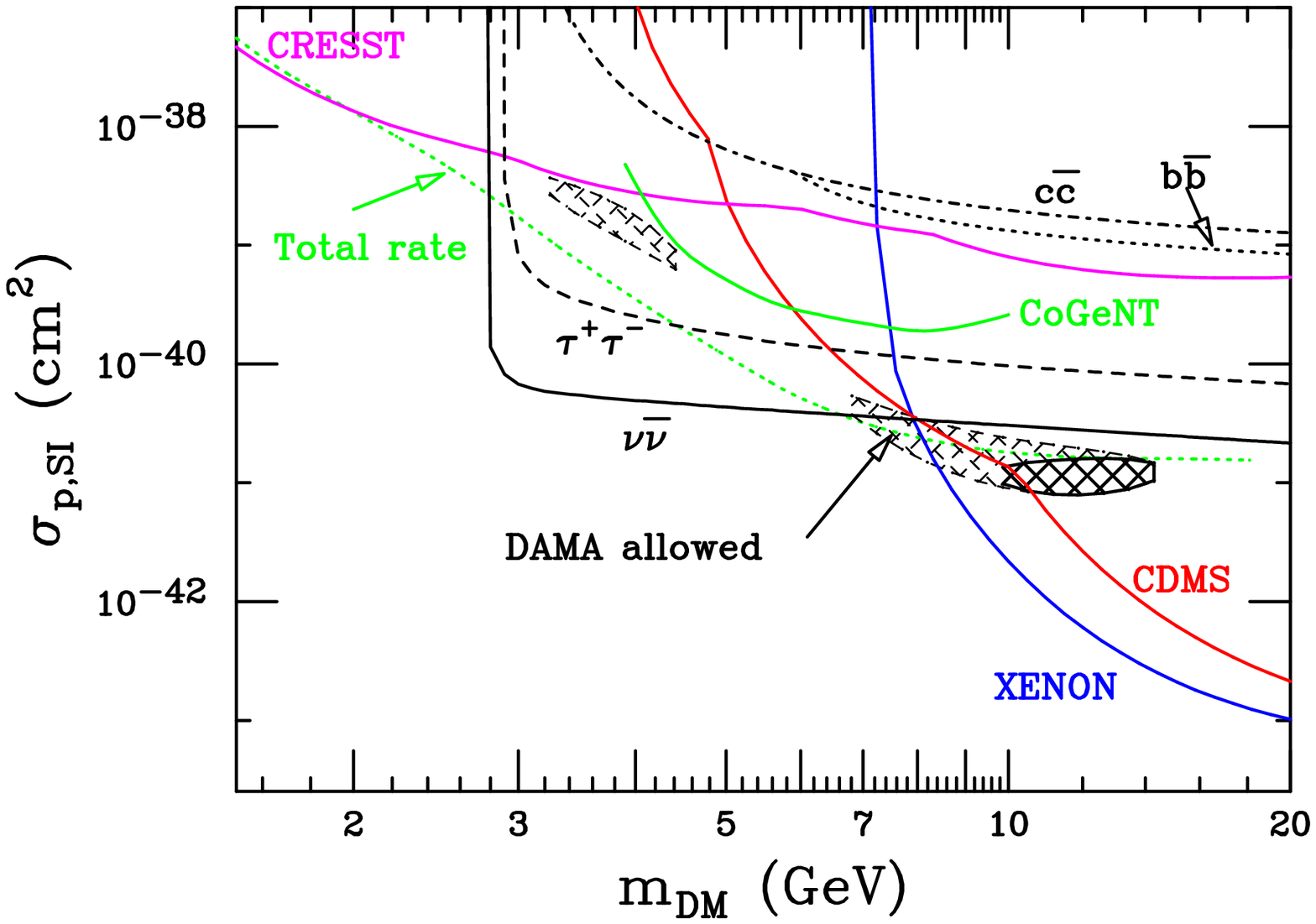}

%\resizebox{11.2cm}{!}{\includegraphics{limitSI.ps}}
%\resizebox{11.2cm}{!}{\includegraphics{limitSI2.ps}}
\caption{The limit on a light WIMP's spin-independent elastic scattering cross section with nuclei from Super-Kamiokande for various choices of dominant annihilation modes. The upper frame contains the DAMA allowed region as calculated in the two-bin analysis of Ref.~\cite{Petriello:2008jj}, while the lower frame uses the full spectral analysis with (dark hatched region) and 
without (light hatched region) the $2-2.5\;{\rm keVee}$ bin.  Also shown are the limits from the CDMS~\cite{cdms}, CRESST~\cite{cresst}, CoGeNT~\cite{cogent}, and XENON\cite{xenon} collaborations.  The dotted green lines are 
the constraints derived by demanding that the predicted total rates predicted by a given WIMP candidate do not exceed those observed by DAMA at the $2\sigma$ level in any energy bin.}
\label{limitSI}
%\end{center}
\end{figure}

\begin{figure}[htbp]
%\begin{center}

 \includegraphics[width=0.61\textwidth,angle=0]{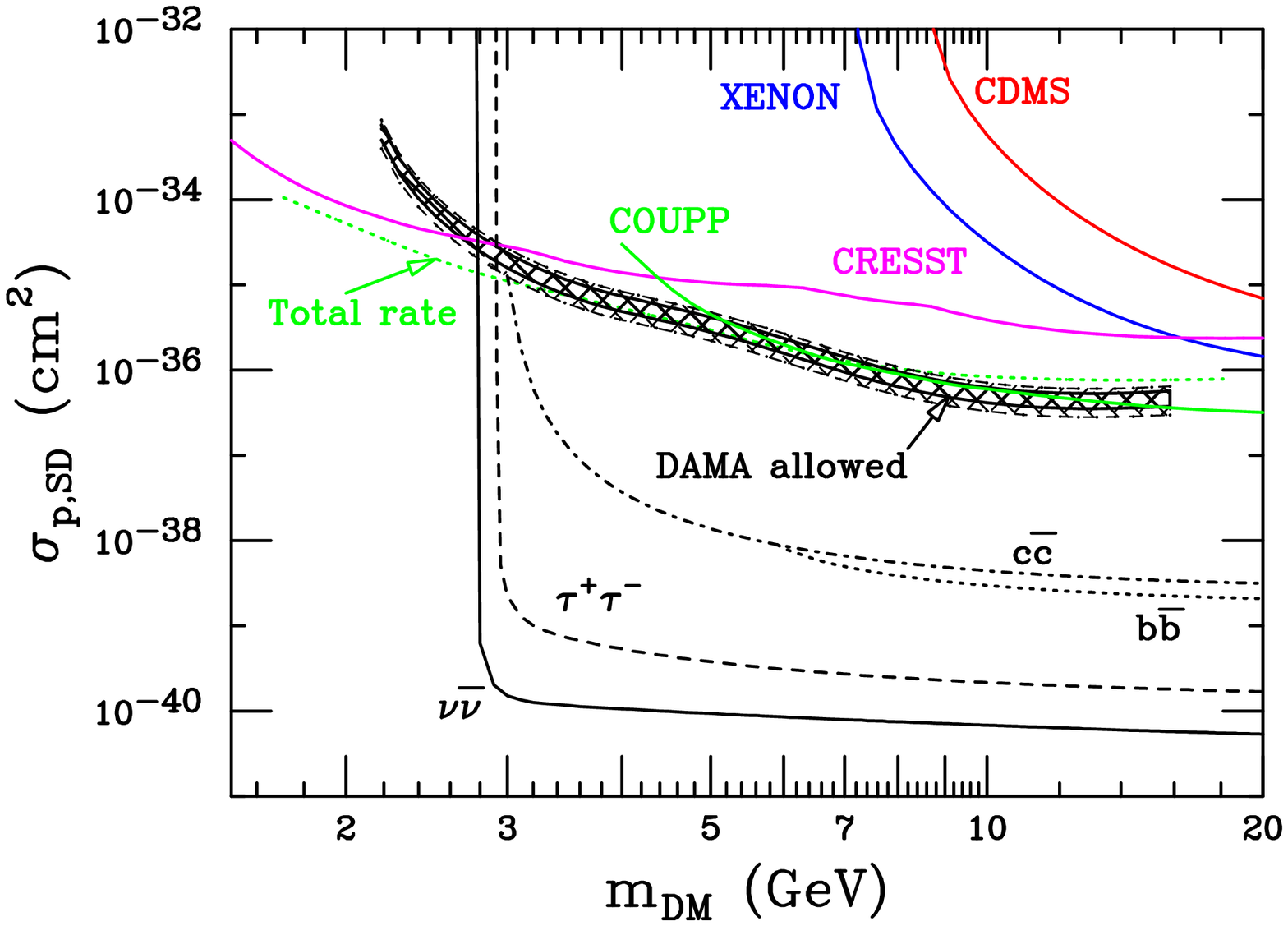}
 \includegraphics[width=0.61\textwidth,angle=0]{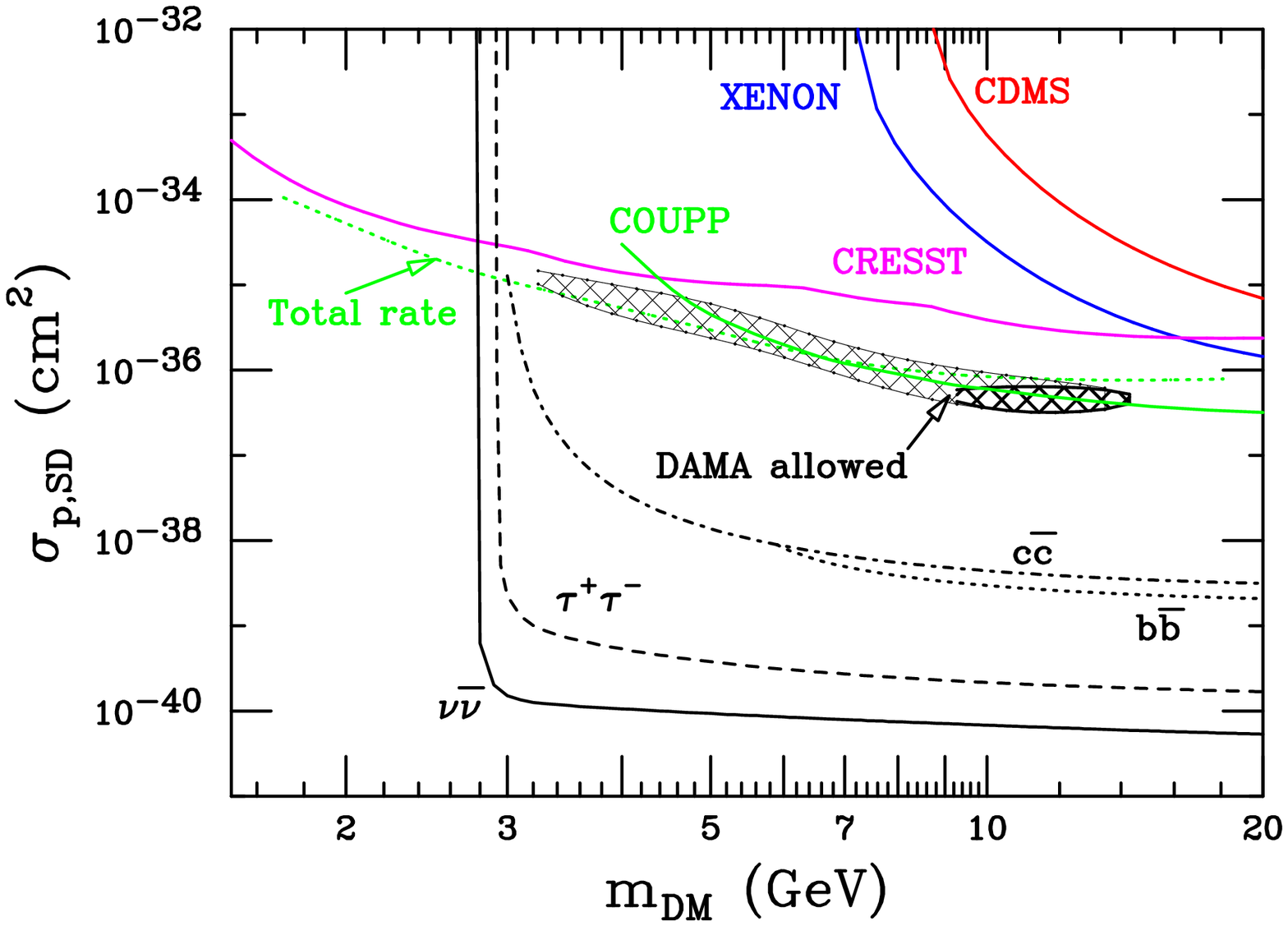}

%\resizebox{11.4cm}{!}{\includegraphics{limitSD.ps}}
%\resizebox{11.4cm}{!}{\includegraphics{limitSD2.ps}}
\caption{The limit on a light WIMP's spin-dependent elastic scattering cross section with nuclei from Super-Kamiokande for various choices of dominant annihilation modes. The upper frame contains the DAMA allowed region as calculated following the two-bin analysis of Ref.~\cite{Petriello:2008jj}, while the lower frame uses the full spectral analysis with (dark hatched region) and 
without (light hatched region) the $2-2.5\;{\rm keVee}$ bin.  Also shown are the limits from the CDMS~\cite{cdms2}, CRESST~\cite{cresst}, XENON~\cite{xenon2} and COUPP~\cite{coupp} collaborations.  The dotted green lines are 
the constraints derived by demanding that the predicted total rates predicted by a given WIMP candidate do not exceed those observed by DAMA at the $2\sigma$ level in any energy bin.}
\label{limitSD}
%\end{center}
\end{figure}

We can translate our predicted rate in Super-Kamiokande to a limit on the WIMP elastic scattering cross section as a function of mass and the dominant annihilation channel. According to the analysis of the Super-Kamiokande collaboration~\cite{superk}, 170 events were observed within an angular window of radius 36$^{\circ}$ centered around the Sun, over 1679.6 live days. Comparing this to the expected rate from atmospheric neutrinos of 185 events, this leads to a 2$\sigma$ upper limit on the contribution from WIMP annihilations of approximately: $[170 + 2 \sqrt{170}]-185 \approx 11$ events. We find a similar result is a somewhat smaller angular window is considered.

In Figs.~\ref{limitSI} and~\ref{limitSD}, we apply this limit and plot our results in the $(\sigma,m_{DM})$ plane for the cases of spin-independent and spin-dependent scattering.  We show both the two-bin analysis procedure of Ref.~\cite{Petriello:2008jj} and the full spectral analysis of the DAMA data both with and without the $2-2.5\;{\rm keVee}$ bin at threshold~\cite{Fairbairn:2008gz}.  
We include in these plots the constraint 
coming from demanding that the total rate observed by the DAMA collaboration not exceed that predicted by a given WIMP mass and cross section.  The form factors and couplings 
required for the calculation of the spin-dependent DAMA allowed
region were obtained from
Refs.~\cite{Ressell:1997kx,Tovey:2000mm}.  We find that if the
WIMP annihilates to neutrinos or taus a significant fraction of
the time, existing data from Super-Kamiokande closes a large
fraction of the DAMA window. In the case of spin-dependent
scattering, WIMPs in the DAMA region which annihilate to
neutrinos, tau leptons, or charm quarks or bottom quarks with any significant probability are ruled out by Super-Kamiokande.  The only possible exception occurs near $m_{\rm DM} \approx 2.6-3.1$ GeV, where evaporation enables the WIMP to evade detection by Super-Kamiokande.

\section{Light Neutralino Dark Matter}
\label{susy}

Within the context of the R-parity conserving minimal supersymmetric standard model (MSSM), a neutralino (if the lightest supersymmetric particle) in the mass range of the DAMA window would very likely be overproduced in the early Universe~\cite{lspmssm}. The main reason for this conclusion is that LEP constraints force the MSSM Higgs bosons, charginos and sfermions to be heavier than $\sim 100$ GeV, in which case they are unable to efficiently mediate the process of neutralino annihilation or to participate in coannihilations. In supersymmetric models with extended Higgs sectors, however, this conclusion can be evaded. In the next-to-ninimal supersymmetric standard model (NMSSM), for example, it has been shown that very light neutralinos ($\sim$1--10 GeV) can be thermally produced in acceptable quantities~\cite{lspnmssm} (see also Ref.~\cite{lspnmssm2}). This is made possible by the exchange of a relatively light pseudoscalar Higgs boson. Intriguingly, if the lightest Higgs scalar is very light and singlet-like, light neutralinos are naturally predicted to possess an elastic scattering cross section near or within the DAMA window. 

Being Majorana fermions, neutralino annihilation to fermions is
chirality suppressed \cite{Goldberg:1983nd}, leading to $\sigma
v \propto m^2_f/m^2_{\chi^0}$. For neutralinos in the mass range
under consideration here, annihilations proceed almost entirely
to combinations of $b\bar{b}$, $\tau^+ \tau^-$ and $c\bar{c}$.  Above the bottom quark threshold, neutralino annihilations mediated via Higgs exchange produce bottom quarks approximately $3 m^2_b/(3 m^2_b + 3 m^2_c + m^2_{\tau}) \sim 90\%$ of the time. For neutralinos near the $b$ mass, however, the phase space for annihilations to $b\bar{b}$ is reduced and the fraction of annihilations producing tau leptons and charm quarks is enhanced. Below the $b\bar{b}$ threshold, at least 40\% of neutralino annihilations proceed to $\tau^+ \tau^-$ (and a considerably larger fraction if the Higgs' couplings to down-type fermions are enhanced by $\tan \beta$).

Comparing this to the results shown in Figs.~\ref{limitSI} and~\ref{limitSD}, we can place constraints on light neutralinos as the source of the DAMA signal.  Considering the case of spin-independent scattering, neutralinos above the $b\bar{b}$ threshold ($4.2 \, \rm{GeV}\, \lsim m_{\chi^0} \lsim 8 \, \rm{GeV}$) annihilate mostly to bottom quarks and are not able to be constrained by Super-Kamiokande beyond the corresponding limits from XENON, CDMS, and CoGeNT. Below the $b\bar{b}$ threshold ($3 \, {\rm{GeV}}\, \lsim m_{\chi^0} \lsim 4.2 \, {\rm{GeV}}$), however, many or most neutralino annihilations produce tau pairs, allowing Super-Kamiokande to exclude this range of neutralino masses. For spin-dependent scattering, the constraints from Super-Kamiokande are even more severe. It is difficult to imagine any neutralino that could generate the observed DAMA signal through spin-dependent scattering without being in conflict with the constraints we have presented in this paper (the possible exception being a neutralino with a $\sim 3$ GeV mass, which could efficiently evaporate in the Sun, thus suppressing the annihilation rate).

\section{Conclusions \label{sec:conc}}

In this paper, we have discussed the indirect detection of light WIMPs in the several GeV range by their capture in the Sun and subsequent annihilation to neutrinos.  Such dark-matter particles can potentially accommodate the annual modulation signal observed by DAMA/LIBRA and the null results of other direct-detection experiments, once the effects of channeling are accounted for. We consider the constraints that can be placed by the Super-Kamiokande experiment on WIMPs in the DAMA window for the cases of prompt dark-matter annihilation to neutrinos, and 
secondary neutrinos produced by decays of tau leptons and charm and bottom quarks.  The constraints found are significant, and impose stringent constraints on the DAMA allowed parameter space, particularly in the low mass end of the DAMA region.  For the case of spin-independent 
elastic scattering of WIMPs with nuclei, dark-matter particles that annihilate to tau leptons or neutrinos a significant fraction of the time are excluded by Super-Kamiokande measurements.  For 
spin-dependent scattering, the constraints are more severe;
any annihilation fraction to neutrinos, tau leptons, or charm or bottom quarks above the $10^{-2}$ level is ruled out.

\acknowledgements{DH is supported by the Fermi Research Alliance, LLC under Contract No.~DE-AC02-07CH11359 with the US Department of Energy and by NASA grant NNX08AH34G.
MK is supported by DoE grant DE-FG03-92-ER40701 and the Gordon and Betty Moore Foundation.  FP and KMZ are supported by  the DOE grant DE-FG02-95ER40896, Outstanding  Junior Investigator Award, by the University of Wisconsin Research Committee
with funds provided by the Wisconsin Alumni Research Foundation, and
by the Alfred P.~Sloan Foundation.  FP and KMZ thank the Aspen Center for Physics, where part of this work was completed.
}

\end{document}